\documentstyle[prb,aps,floats,epsf,rotating]{revtex}
\begin{document}


\twocolumn[\hsize\textwidth\columnwidth\hsize\csname@twocolumnfalse%
\endcsname
\title{Orbital order out of spin disorder: How to measure the orbital gap}
\author{G.~Khaliullin\protect\cite{K*}}
\address{Max-Planck-Institut f\"ur Physik komplexer Systeme, 
N\"othnitzer Strasse 38, D-01187 Dresden, Germany\\
and Max-Planck-Institut f\"ur Festk\"orperforschung, Heisenbergstrasse 1,
D-70569 Stuttgart, Germany}
\author{R.~Kilian}
\address{Max-Planck-Institut f\"ur Physik komplexer Systeme, 
N\"othnitzer Strasse 38, D-01187 Dresden, Germany}
\date{February 3, 1999}
\maketitle


\begin{abstract}
The interplay between spin and orbital degrees of freedom in the Mott-Hubbard 
insulator is studied by considering an orbitally degenerate superexchange
model. We argue that orbital order and the orbital excitation gap in this
model are generated through the order-from-disorder mechanism known
previously from frustrated spin models. We propose that the orbital gap
should show up indirectly in the dynamical spin structure factor;
it can therefore be measured using the conventional inelastic neutron
scattering method.
\end{abstract}
\draft
]


The recent renaissance in the study of transition metal oxides has 
emphasized the important role being played by the orbital (pseudo) 
degeneracy inherent to perovskite lattices. First of all,
the type of spin structure and the character of spin excitations crucially
depend on the orientation of occupied orbitals \cite{GOO55,KAN59,KUG82}. 
Second, the excitations in the orbital sector get coupled to the other
degrees of freedom (electronic, lattice, spin) and might therefore
strongly modify their excitation spectra. It was also suggested 
recently \cite{ISH97a,KIL98} that low-energy orbital fluctuations
are responsible for the highly correlated metallic state of 
ferromagnetic manganites. 

Apparently, orbital order and orbital fluctuations deserve
for careful theoretical and experimental study. However, the
orbital excitation itself is spin- and chargeless and can therefore be
detected only indirectly due to its coupling to other
types of excitations. For instance, Ishihara {\it et al.} \cite{ISH98}
have recently discussed the possibility to detect orbital excitations
by means of the anomalous x-ray scattering method. In this paper
we propose the idea of detecting the orbital excitation in a 
conventional inelastic neutron scattering experiment. That is, due
to the inherent coupling of orbitals to the spin, a magnon and a
single orbital wave can be excited by neutrons. 
Below, we illustrate this idea by considering an $e_g$ orbitally
degenerate antiferromagnetic model. 

Physically realistic spin-orbital models \cite{KUG82,ISH97b} are usually
rather complicated. To make the discussion more transparent we
consider a simplified version of the Kugel-Khomskii model 
\cite{KUG82,FEI97}
\begin{eqnarray}
H &=& \sum_{\langle i j \rangle } \Big[
4(\vec S_{i}\vec S_{j})
(\tau_{i}^{\alpha}-\frac{1}{2})
(\tau_{j}^{\alpha}-\frac{1}{2})\nonumber\\
&&+(\tau_{i}^{\alpha}+\frac{1}{2})
 (\tau_{j}^{\alpha}+\frac{1}{2})-1 \Big]
\label{HAM}
\end{eqnarray}
describing the superexchange process in a Mott-Hubbard insulator
with $e_g$ degeneracy, where the Hund's splitting in the intermediate
state has been neglected. 
Here, $\vec S_i$ is the spin $1/2$ operator, while operators $\tau_{i}^{\alpha}$ 
act in the orbital subspace with basis vectors $(1,0)$, 
$(0,1)$ corresponding to the $e_g(x^2-y^2)\sim \mid x \rangle$ 
and $e_g(3z^2-r^2)\sim \mid z \rangle$ orbital states, respectively.
The structure of $\tau_i^{\alpha}$ depends on the index $\alpha$ which
specifies the orientation of the bond $\langle ij \rangle$ relative to
the cubic axes $a,b,c$:
\begin{equation}
\tau_{i}^{a(b)}=\frac{1}{4}
(-\sigma_{i}^{z}\pm \sqrt{3}\sigma_{i}^{x}),\;\;\;\;\;
\tau_{i}^{c}=\frac{1}{2}\sigma_{i}^{z},
\end{equation}
where $\sigma^z$ and $\sigma^x$ are Pauli matrices. The physical
meaning of the $\tau$ operators is to describe the fluctuations
of exchange bonds due to orbital dynamics.
On the cubic lattice, Eq.~(\ref{HAM}) can be rewritten in 
the equivalent form:
\begin{eqnarray}
H &=& -3+\sum_{\langle i,j \rangle } \hat J^{ij}_\alpha
( \vec S_i \vec S_j + \frac{1}{4}),
\label{HAM2}\\
\hat J^{ij}_{\alpha} &=& 4 \tau_{i}^{\alpha} \tau_{j}^{\alpha} -
2(\tau_{i}^{\alpha}+  \tau_{j}^{\alpha} )+1\nonumber.
\end{eqnarray}

The main feature of this model is the strong interplay between
spin and orbital degrees of freedom which is suggested by the
very form of Hamiltonian~(\ref{HAM2}). In fact, the Kugel-Khomskii
model contains rather nontrivial physics: That is, the classical
N\'eel state in Eq.~(\ref{HAM2})
($\langle \vec{S}_i \vec{S}_j \rangle = -1/4$) is
infinitely degenerate due to the presence of the orbital sector,
and this extra degeneracy must be lifted by some mechanism.
According to a recent proposal by Feiner, Ole\'s, and Zaanen \cite{FEI97}, 
this orbital frustration problem is likely solved by the formation of a  
RVB-type spin-singlet ground state. Technically, their suggestion is based 
on the observation that spin-orbit coupling results in a new,
composite (simultaneous spin and orbital flip) excitation. This
mode is found to be soft in certain directions in momentum
space, leading to one-dimensional fluctuations, thus destroying
the magnetic order completely. On the contrary, Khaliullin and
Oudovenko \cite{KHA97} have found the orbital flip excitation to
be gapped, and concluded that the orbitally ordered quasi
one-dimensional quantum N\'eel state is the proper low-temperature
fixed point of model~(\ref{HAM2}). 

The purpose of our paper is twofold. First, we reexamine the 
solution of Feiner {\it et al.} \cite{FEI97} and show that
the soft spin-orbital mode acquires in fact a finite gap through
Villain's order-from-disorder mechanism \cite{TSV95}. 
Basically, this gap is determined by the orbital excitation energy. 
Second, we calculate the spectral weight of the composite
spin-orbital
excitation in the dynamical spin structure factor, suggesting
the possibility to measure the orbital excitation gap
indirectly by a conventional neutron scattering experiment.

\begin{figure}
\centering
\setlength{\unitlength}{0.9\linewidth}
\begin{picture}(1,1)
\put(0,0){
\epsfysize=\unitlength
\epsffile{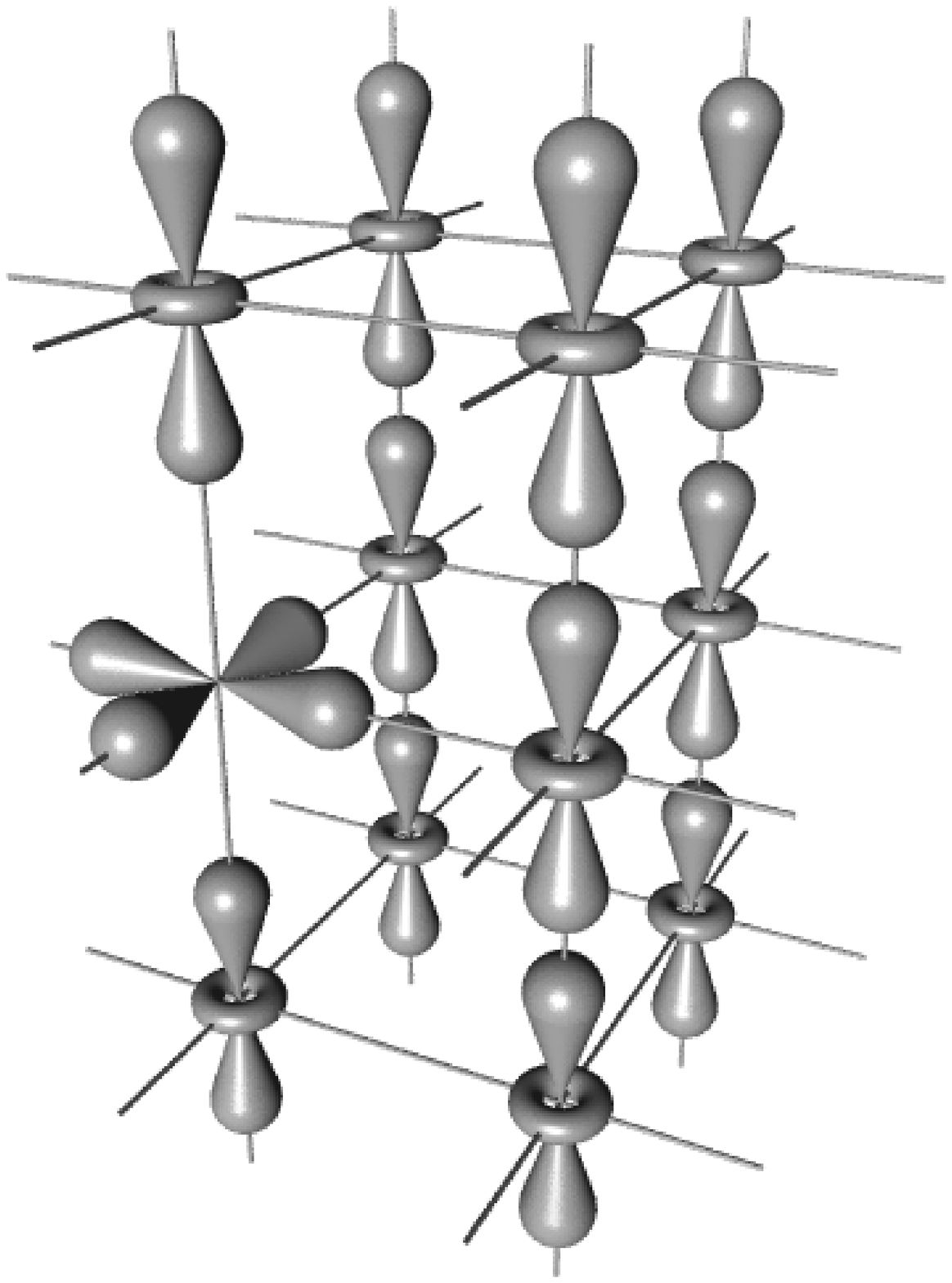}}
\put(0.12,0.53){
\epsfysize=0.15\unitlength
\epsffile{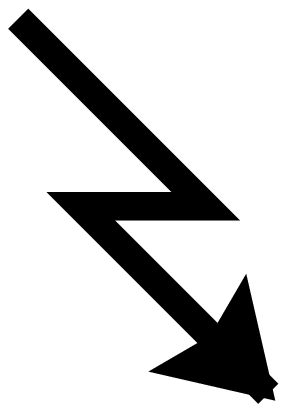}}
\end{picture}
\caption{$|3z^2-r^2\rangle$ orbital order which leads to weakly
coupled AF spin chains ($J_c=4$, $J_{\perp} = 1/4$).
As discussed in Ref.\ \protect\onlinecite{KHA97}, this type of orbital 
ordering provides the largest energy gain due to quantum spin 
fluctuations. An orbital flip (indicated by an arrow)
modulates the strength of the neighboring exchange bonds,
breaking the $c$ chain. In
the classical N\'eel state, orbital excitations cost no energy. 
However, strong quasi one-dimensional spin fluctuations
stabilize this structure by opening the orbital gap through
a order-from-disorder mechanism.}
\label{FIG:ORB}
\end{figure}

To study model~(\ref{HAM}) we follow the same scheme and 
method used by Feiner {\it et al.} \cite{FEI97}. Namely, 
i) we start with the assumption of long-range staggered spin order 
and an uniform $z$-type orbital ordering  (see Fig.\ \ref{FIG:ORB});
ii) next we calculate the transverse spin-fluctuation spectrum by using 
the equation-of-motion method, and
calculate the quantum corrections to the magnetic order parameter.
We are motivated to use this method here for the following reason: 
Employing a conventional diagrammatic  technique it was argued
in Ref.\ \onlinecite{KHA97} that the orbital degeneracy of the
classical state is removed by quantum spin fluctuations which
generate orbital order and spontaneously  break the cubic symmetry.
The interesting question is then whether and how this order from 
disorder phenomena manifests itself in the equation of motion
ansatz of Ref.\ \onlinecite{FEI97}.

The quantity to be calculated is the Green's function 
$G_{i,i'} = \langle\langle S_i^+|S_{i'}^-\rangle\rangle$, from which 
the dynamical spin structure factor as well as the reduction of the spin 
order parameter can be deduced. It obeys the following equation of 
motion: 
\begin{equation}
\omega G_{i,i'} = \langle[S_i^+,S_{i'}^-]\rangle + \langle\langle
[S_i^+,H]|S_{i'}^-\rangle\rangle.
\end{equation}
Within a conventional spin-wave approximation, this equation reads as
\begin{eqnarray}
\left(\omega-\lambda m_i\right) G_{i,i'} &=& \delta_{i,i'} +
2m_i\sum_{j_c} G_{j,i'} + \frac{1}{8}m_i\sum_{j_{\perp}} G_{j,i'}\nonumber\\
&&+\frac{\sqrt{3}}{8} m_i \sum_{j_{\perp}} \alpha_{ij} D_{j,i'}.
\label{SWA}
\end{eqnarray}
The following notations are introduced here: $m_i=1$ if site $i$ belongs to
sublattice $A$ with up-spin orientation, and $m_i=-1$ otherwise.
The summations over $j_c$ and $j_{\perp}$ are over nearest neighbors of site
$i$ in the $c$-chain and perpendicular directions, respectively. The factor
$\lambda=J_c+2J_{\perp} = 9/2$ and $\alpha_{ij}=|R_{ij}^x|-|R_{ij}^y|$. 
Distinct from a Heisenberg model, a new Green's function $D_{j,i'} =
\langle\langle K_{jj}^x|S^-_{i'} \rangle\rangle$ appears
in the above equation. The operator $K_{jj'}^x =
\sigma_j^xS_{j'}^+$ describes a simultaneous orbital-spin excitation which
couples to a single spin flip. Following Ref.\ \onlinecite{FEI97}, we discarded
in Eq.\ (\ref{SWA}) the term $K_{j\ne j'}$, neglecting the correlation between
orbital and spin excitations from different chains.

The dynamics of the composite excitations 
$K_{jj}^{\pm} =\sigma_j^{\pm}S_j^+$ is described by the equation
\begin{eqnarray}
\left(\omega+3m_i\mp\frac{3}{2}\right) K_{jj}^{\pm} &=&
-\frac{3}{16} (m_i\pm 1) \sum_{l_{\perp}} \left(
K_{ll}^x + \frac{\alpha_{jl}}{\sqrt{3}} S_l^+\right)\nonumber\\ 
&&+ L_j^{\pm},
\label{EQK}
\end{eqnarray}
where the last term is
\begin{equation}
L_j^{\pm} = \sum_{l_c} \left\{ 2K_{jl}^{\pm} \delta S_j^z -2K_{jj}^{\pm}
\delta S^z_l - (m_i\mp 1) K_{jl}^{\pm}\right\} .
\label{EQL}
\end{equation}
Here, site $l$ is the nearest neighbor of site $j$, and $\delta S^z_j$ is the
fluctuating part of the spin operator $S^z_j$. The $L^{\pm}_j$ term
in Eqs.\ (\ref{EQK}) and (\ref{EQL}) accounts for the correlations within
the $c$ chain. If one neglects this term (as Feiner {\it et al.} \cite{FEI97}
did), one obtains from Eqs.\ (\ref{SWA}) and (\ref{EQK}) a soft mode
mentioned above, which results finally in a breakdown of the assumed
spin-ordered state. We do not accept this approximation. It is the
intra-chain quantum fluctuations represented by the $L^{\pm}_j$ term 
that play a crucial role by making the orbital excitations $\sigma^{\pm}$
(and consequently $\sigma^{\pm}S^+$) to acquire a finite mass gap.
The underlying physics can easily be observed in the following way:
We write the equations of motion for the operators in Eq.\
(\ref{EQL}) treating the bond $(jl)$ exactly while considering the
remaining bonds of the $c$ chain in a static N\'eel approximation. This results in
\begin{eqnarray}
\left[\omega-(m_i\pm 3)\right] K^{\pm}_{jl} \delta S^z_j &=&
(1\pm m_i) \left( K_{jj}^{\pm} \pm K_{jj}^{\pm} \delta S_l^z\right),\nonumber\\
(\omega\mp 2) K_{jj}^{\pm} \delta S^z_l &=& (1\mp m_i) \left(
K_{jl}^{\pm} \pm K_{jl}^{\pm} \delta S_j^z\right),\nonumber\\ 
\left[ \omega -(3m_i\pm 1)\right] K^{\pm}_{jl} &=& 2 K_{jj}^{\pm}
\delta S^z_l -2 K_{jl}^{\pm} \delta S^z_j\nonumber\\
&&+ (m_i\pm 1) K_{jj}^{\pm}.
\label{BND}
\end{eqnarray}
Many terms in these equations in fact drop out due to $|m_i|=1$, and 
Eqs.\ (\ref{BND}) together with Eq.\ (\ref{EQL}) lead to a very simple 
result. Namely, 
\begin{equation}
L^+_{j\in B} = \frac{8}{\omega-4} K^+_{jj}, \quad 
L^-_{j\in A} = \frac{8}{\omega+4} K^-_{jj},
\end{equation}
and $L^-_{j\in B} = 0$, $L^+_{j\in A}=0$. Now, Eq.\ (\ref{EQK})
can be written as follows (note $K^x=K^+ + K^-$):
\begin{eqnarray}
(\omega+\frac{3}{2}m_i)K^x_{jj} &=& -\frac{3}{8} m_i \sum_{l_{\perp}}
\left(K^x_{ll} + \frac{\alpha_{jl}}{\sqrt{3}} S_l^+\right)\nonumber\\ 
&&+\Sigma_j K_{jj}^x.
\label{SIG}
\end{eqnarray}
The function $\Sigma_j=8/(4m_j+\omega)$ is interpreted as a self-energy
correction to the composite excitation $\sigma_j^xS_j^+$ due to quantum
fluctuations about the N\'eel state. We think that the large $\omega$
behavior of $\Sigma$ is not quite reliable since short-time fluctuations
of the environment of the bond $(jl)$ were neglected in the above crude
derivation. We therefore take the low-energy limit, $\Sigma_j \approx
2m_j-\frac{1}{2}\omega$, which is of main interest. Physically, the 
self-energy is mainly due to the orbital flip excitation (see Fig.\ 
\ref{FIG:ORB}) which acquires a finite energy as strong correlations within 
the $c$ chains are explicitly taken into account. 

\begin{figure}
\noindent
\centering
\begin{turn}{-90}
\epsfysize=0.8\linewidth
\epsffile{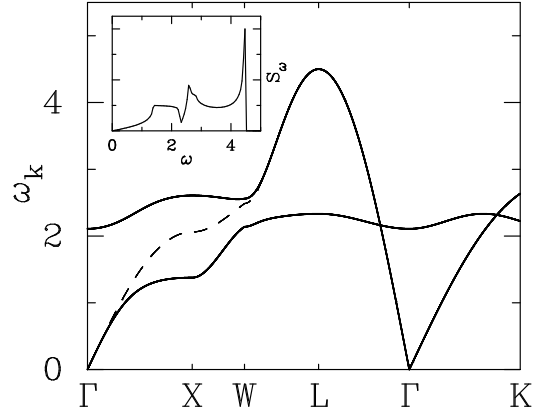}
\end{turn}\\[6pt]
\caption{Transverse spin excitation spectrum along the direction
$\Gamma$ - $\text{X}(\pi,0,0)$ - $\text{W}(\pi,\frac{\pi}{2},0)$ - $\text{L}
(\frac{\pi}{2},\frac{\pi}{2},\frac{\pi}{2})$ - $\Gamma$ - 
$\text{K}(\frac{3\pi}{4},\frac{3\pi}{4},0)$ in the Brillouin zone. To the right
from L, the spin-orbit coupling vanishes and excitations are of either
pure spin flip or of simultaneous spin-orbital flip character. Solid
lines: present theory. Dashed line: spin-wave dispersion calculated 
with $J_c=4$, $J_{\perp}=1/4$ and neglecting orbital fluctuations.
Inset: on-site dynamic structure factor; the pseudogap seen about
$\omega \approx 2$ is the manifestation of the orbital gap.}
\label{FIG:SP1}
\end{figure}

\begin{figure}
\noindent
\centering
\begin{turn}{-90}
\epsfysize=0.8\linewidth
\epsffile{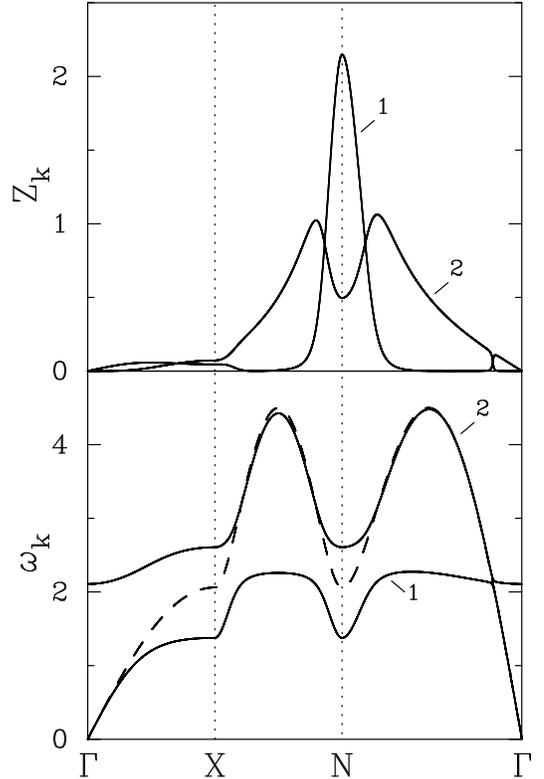}
\end{turn}\\[6pt]
\caption{Transverse spin excitation spectrum (lower panel) and the
corresponding
spectral weights in the dynamical structure factor (upper panel) along
the $\Gamma$ - $\text{X}(\pi,0,0)$ - $\text{N}(\pi,0,\pi)$ - $\Gamma$
direction. Dashed line: spin-wave spectrum calculated neglecting
spin-orbit coupling.}
\label{FIG:SP2}
\end{figure}

We now denote $G_s = G_{AA}$, $D_s = D_{AA}$ if the index $s=1$,
and $G_s=G_{BA}$, $D_s=D_{BA}$ if $s=-1$. In momentum space,
Eqs.\ (\ref{SWA}) and (\ref{SIG}) lead to 
\begin{eqnarray}
(s\omega-\lambda) G_s - \lambda \gamma_k G_{-s} +V_k D_{-s} &=& 
\delta_{s,1},\nonumber\\
\left[s\omega - \tau - \Sigma(s\omega)\right] D_s - \tau \gamma_{\perp} 
D_{-s} + V_k G_{-s} &=& 0.
\label{LAM}
\end{eqnarray}
Here, $\Sigma(\omega) = 2-\omega/2$,
$\tau=3/2$, $\gamma_k=(8c_z+\gamma_{\perp})/9$, $\gamma_{\perp} = 
(c_x+c_y)/2$, $ V_k = \sqrt{3}(c_x-c_y)/2$, and $c_{\alpha} = \cos
k_{\alpha}$.
Eqs.\ (\ref{LAM}) can easily be solved to give two branches $\omega_1(k)$
and $\omega_2(k)$ for the spin excitations and their relative weights in a 
dynamic spin structure factor. The results are shown in Figs.\ \ref{FIG:SP1}
and \ref{FIG:SP2} for certain directions in momentum space.

It should be noticed first that there is only one gapless mode at the
$\Gamma$ point. This is a plausible result since the original model
has a continuous symmetry only in spin subspace; the second
branch related to the breaking of the discrete orbital symmetry
is expected to have a finite gap. We conclude that the soft mode
discussed in Ref.\ \onlinecite{FEI97} is an artefact of the approximation
which neglects the crucial effect of quantum spin fluctuations on
the composite excitation $\sigma^x S^+$. With regards to the
low energy-momentum limit, the model (\ref{HAM}) behaves
as a conventional Heisenberg system. However, the underlying
anisotropic orbital ordering and the presence of orbital 
fluctuations strongly enhance quantum effects and reduce the
order parameter. We have calculated $\langle S^z\rangle = 0.23$
which is consistent with Ref.\ \onlinecite{KHA97}. 

Second, at the magnetic zone boundaries the spin-orbit coupling results 
in quite visible deviations of the main magnon branch from that of an 
anisotropic Heisenberg model (see Figs.\ \ref{FIG:SP1} and \ref{FIG:SP2}).
This effect, which is due to the modulation of exchange bonds by orbital
fluctuations, is expected to be a generic feature of spin-orbital
models. In the present model, the zone boundary effect is 
related to the particular momentum dependence of the coupling
constant $V_k$ in Eq.\ (\ref{LAM}).
Experimentally, anomalous zone boundary magnon softening
has been observed in ferromagnetic manganites \cite{HWA98}.
We have recently suggested \cite{KHA99} that this effect originates from
the modulation of double-exchange bonds by orbital fluctuations. 

Because of the finite hybridization of the excitations $S^+$ and
$\sigma^x S^+$ (note that orbital pseudospin is not a conserved
quantity), a conventional inelastic neutron scattering experiment
might provide information on orbital excitation energies as well.
In Fig.\ \ref{FIG:SP2}, we plot the relative spectral weight of two
modes in the dynamical structure factor. A strong mixture of single
spin $S^+$ and $\sigma^x S^+$ excitations and the ``level
repulsion'' effect take place at energies where these excitations
meet each other. In the present model, this results in an additional peak
which has substantial weight at certain momenta. 

One comment is necessary at this point: The present approach (as well
as the one of Ref.\ \onlinecite{FEI97}) implicitly assumes the formation 
of a bound state of orbital and spin excitations (a tightly bound composite
excitation). However, it might well be possible that the bound state
decays into an independent orbital wave and magnons. In this case,
one should expect instead of a well defined additional peak a softening
and damping of magnons due to the coupling to the orbital-magnon 
excitation continuum. In fact, the latter picture is described in Ref.\
\onlinecite{KHA97}, which treats the spin-orbit coupling
perturbatively. In some low-dimensional spin-orbital models 
the bound state might exist \cite{BRI98}. The problem of a bound
state versus particle-hole continuum in a Kugel-Khomskii model
remains open at present. In either case, with regards to the low-energy
behavior both approaches~-- the present one as well as that of Ref.\ 
\onlinecite{KHA97}~-- are fully consistent with each other since the 
essence of the model, that is the orbital-order-out-of-spin-disorder
mechanism, is equally taken into account. 

To conclude, we have studied a toy model providing a strong interplay
between spin and orbital excitations. In the general
context of transition metal oxides, this work suggests that the orbital
fluctuations lead to a softening and damping of the zone-boundary
magnons, and may even result in an additional structure in the
spin-response function of these compounds.


\end{document}